\documentclass{article}
\usepackage{amssymb}
\usepackage{graphicx}
\begin{document}
\begin{center}
{\Large\bf LATE TIME ACCELERATED\\[5pt]
BRANS-DICKE PRESSURELESS SOLUTIONS\\[5pt]
AND THE SUPERNOVAE TYPE Ia DATA\\[5pt]}
\medskip

{\bf J.C. Fabris\footnote{e-mail: fabris@cce.ufes.br}, S.V.B.
Gon\c{c}alves\footnote{e-mail: sergio@cce.ufes.br} and R. de S\'a
Ribeiro\footnote{e-mail: ribeiro@cce.ufes.br}} \medskip

Departamento de F\'{\i}sica, Universidade Federal do Esp\'{\i}rito
Santo,\\
Vit\'oria, 29060-900, Esp\'{\i}rito Santo, Brazil
\medskip

\end{center}

\begin{abstract}
Cosmological solutions for a pressureless fluid in the Brans-Dicke
theory exhibit asymptotical accelerated phase for some range of
values of the parameter $\omega$, interpolating a matter dominated
phase and an inflationary phase. The effective gravitational
coupling is negative. We test this model against the supernovae
type Ia data. The fitting of the observational data is slightly
better than that obtained from the $\Lambda CDM$ model. We make
some speculations on how to reconcile the negative gravitational
coupling in large scale with the local tests. Some considerations
on the structure formation problem in this context are also made.
\vspace{0.7cm}

PACS number(s): 04.20.Cv., 04.20.Me
\end{abstract}

\section{Introduction}

The accumulation of supernovae type Ia data seems to confirm that
the universe is today in an accelerated phase
\cite{riess,perlmu1}. The first indications of this accelerated
phase came from the analysis of about $50$ supernovae. Today, the
data are approaching 300 supernovae \cite{tonry,riessbis}, and
those first conclusions, for the moment, are confirmed. The
theoretical explanation of this surprising result is one of the
most important challenge in cosmology. In order to obtain an
accelerated phase, a repulsive gravitational effect must be
obtained. Generally, this is achieved by introducing an exotic
fluid which exhibits a negative pressure today, called dark energy
(for a recent review, see reference \cite{sahni}). The most
natural candidate is a cosmological constant,  since it is an
inevitable consequence of considering quantum fields in a curved
space-time. But, the cosmological constant proposal faces a major
difficulty: the theoretical predicted value surpass the value
induced by observation by 120 order of magnitudes. This
discrepancy can be alleviated, but not solved, by considering, for
example, supersymmetric theories. Moreover, the introducing of a
cosmological constant today does not explain the coincidence
problem, the fact that the universe begun to accelerated very
recently, after the completion of formation of local structures
like galaxies and clusters of galaxies.
\par
Many other alternatives to the cosmological constant have been
presented in the literature. In the quintessence program, a
self-interacting scalar field is considered \cite{caldwell,wang}.
The potential for the scalar field is such that initially the
pressure is positive, in order to allow the formation of local
structures, becoming latter negative, driving the acceleration of
the universe. The coincidence problem may be addressed in this
case. However, fine tuning is required. However, some more natural
potentials have been determined using supergravity theories
\cite{brax}. Other dark energy models appear in the literature:
k-essence \cite{mukhanov}, Chaplygin gas \cite{pasquier,fabris},
viscous fluid \cite{rose}, etc. Here, we explore another
possibility: the acceleration phase could be generated by a
non-minimal coupling between a scalar field and the gravitational
term.
\par
The paradigm of a scalar-tensor theory with non-minimal coupling is the Brans-Dicke theory,
which is characterized by the Lagrangian
\begin{equation}
\label{lagran}
L = \sqrt{- g}\biggr[\phi R - \omega\frac{\phi_{;\rho}\phi^{;\rho}}{\phi}\biggl] + L_m \quad ,
\end{equation}
where $\omega$ is a coupling parameter  and $L_m$ is a matter
Lagrangian. Local tests at level of the solar system and stellar
binary systems, restrict $\omega$ to be larger than around $1,000$
\cite{will}. This makes the predictions of the Brans-Dicke theory
almost indistinguishable of those coming from general relativity.
The scalar field $\phi$ is related to the inverse of the
gravitational coupling $G$. Some recent interesting proposals
address the possibility that the gravitational coupling can depend
on the scale, having different value at local and at cosmological
scales \cite{mannheim,merab1}. If such scale-dependent
gravitational coupling can be satisfactorily implemented, it is
possible to have smaller values for the parameter $\omega$ at
cosmological scales, obtaining scenarios that differ substantially
from those obtained from general relativity, while agreement with
local tests is preserved. Studies of quantum effects in
gravitational theories, and their consequence for the behaviour of
the gravitational coupling, have strengthened those proposals
\cite{reuter,shapiro}.
\par
In reference \cite{gurevich}, general cosmological solutions for a
flat universe in the Brans-Dicke theory were determined. For a
pressureless fluid, such solutions can interpolate a decelerating
phase with an accelerated phase. This is an example of how
cosmological models in Brans-Dicke theory may differ substantially
from the standard cosmological model: using general relativity
such interpolation can just be obtained through introduction of an
exotic fluid. However,  the price to pay is not negligible: the
Brans-Dicke parameter must takes values in the interval $- 3/2 <
\omega < - 4/3$, implying that the effective gravitational
coupling is negative. The possibility of a scale-dependent
gravitational coupling may render, however, such scenarios
attractive.
\par
These solutions with a late accelerated phase were studied in
reference \cite{rose1}. Here we make a step further: a comparison
with the supernovae type Ia data is made, using the "gold sample".
This Brans-Dicke late accelerated model gives better agreement
with the observational data than the $\Lambda CDM$ model. The best
fitting for the Brans-Dicke parameter and the Hubble parameter $h$
are obtained, indicating $\omega \approx -1.5$ and $h \approx
0.6$. The approach here differs from others existing in the
literature which studies accelerated models in Brans-Dicke theory
\cite{orfeu,quiros,amendola,kim,saulo,davidson,arik,banerjee} by
the fact that no other ingredient is introduced, like a potential
term for the scalar field. But, the model requires an effective
mechanism that leads to a repulsive gravitational coupling at
cosmological scales, while keeping it attractive at local scales.
\par
This article is organized as follows. In next section, the
pressureless Brans-Dicke cosmological model is revised. In section
$3$, the comparison with the supernovae type Ia "gold sample" is
made. In section $4$ we present our conclusions.

\section{Late accelerated phase in the Brans-Dicke cosmological model}

Let us consider a flat Brans-Dicke cosmology, with the universe
filled with pressureless matter. The equations of motion resulting
from the Lagrangian (\ref{lagran}) are:
\begin{eqnarray}
\label{em1} 3\biggr(\frac{\dot a}{a}\biggl)^2 &=&
\frac{8\pi\rho}{\phi} +
\frac{\omega}{2}\biggr(\frac{\dot\phi}{\phi}\biggl)
- 3\frac{\dot a}{a}\frac{\dot\phi}{\phi} \quad ,\\
\label{em2}
\ddot\phi + 3\frac{\dot a}{a}\dot\phi &=& \frac{8\pi\rho}{3 + 2\omega} \quad ,\\
\label{em3} \dot\rho + 3\frac{\dot a}{a}\rho &=& 0 \quad .
\end{eqnarray}
The last equation leads to $\rho = \rho_0a^{-3}$. Inserting this
relation in (\ref{em2}), we find the first integral, $\dot\phi =
\frac{8\pi\rho t}{3 + 2\omega}C$, where $C$ is a constant.
\par
Following \cite{weinberg}, we can define an auxiliary function
$u$, satisfying the relation
\begin{equation}
\label{rel} \frac{\dot u}{u} = -3\frac{\dot a}{a} + \frac{2}{t} -
\frac{u}{t}
\end{equation}
which, in view of the equations of motion, results in the integral
relation
\begin{equation}
\label{int} \int\frac{du}{u[u + 4 \pm s\sqrt{u^2 + 4u}]} = 2\ln (t
- t_c)\quad,
\end{equation}
where $s = 3\sqrt{1 + \frac{2}{3}\omega}$. This integral relation
has three critical points $u = 0$, $u = - 4$ and $u = \frac{2}{4 +
3\omega}$. The first one gives non physical results, while the
second one leads to $a \propto t^2$ for any value of $\omega$. For
this last case, $\frac{8\pi\rho}{\phi} = - 4(3 + 2\omega)$.
However, the fact that the evolution of the scale factor does not
depend on $\omega$ makes this solution not very interesting.
\par
The third critical point of (\ref{int}) leads to a particular
solution in terms of power law function (choosing $t_c = 0$):
\begin{equation}
\label{ps} a = a_0t^\frac{2 + 2\omega}{4 + 3\omega} \quad , \quad
\phi = \phi_0t^\frac{2}{4 + 3\omega} \quad .
\end{equation}
A more general solution may be obtained through integration of
(\ref{int}). This has been done in \cite{gurevich}, resulting in
the general flat solutions
\begin{eqnarray}
\label{g1} a &=& a_0(t - t_-)^\frac{1 + \omega \pm \sqrt{1 +
\frac{2}{3}\omega}}{4 + 3\omega}(t - t_+)^\frac{1 + \omega
\mp \sqrt{1 + \frac{2}{3}\omega}}{4 + 3\omega} \quad , \\
\label{g2} \phi &=& \phi_0(t - t_-)^\frac{1 \mp 3\sqrt{1 +
\frac{2}{3}\omega}}{4 + 3\omega} (t - t_+)^\frac{1 \pm 3\sqrt{1 +
\frac{2}{3}\omega}}{4 + 3\omega} \quad ,
\end{eqnarray}
where $t_\pm = t_c + A(s \pm 1)$, $A$ being another integration
constant. Since $t_+ > t_-$ , $t = t_+$ may be identified with the
initial time.
\par
The solutions presented above have the following properties. For
$t \rightarrow t_+$, they reduce to
\begin{eqnarray}
\label{g3} a &=& a_0t^\frac{1 + \omega
\mp \sqrt{1 + \frac{2}{3}\omega}}{4 + 3\omega} \quad , \\
\label{g4} \phi &=& \phi_0t^\frac{1 \pm 3\sqrt{1 +
\frac{2}{3}\omega}}{4 + 3\omega} \quad ,
\end{eqnarray}
while for $t >> t_+$, they take the form
\begin{equation}
a = a_0t^\frac{2 + 2\omega}{4 + 3\omega} \quad , \quad \phi = \phi_0t^\frac{2}{4 + 3\omega}
\quad .
\end{equation}
\par
Concerning the possible scenarios, the solutions have the
following properties:
\begin{enumerate}
\item For $\omega > -\frac{4}{3}$: The universe has a decelerated
expansion during all its evolution. \item For $\omega < -
\frac{3}{2} < \omega < - \frac{4}{3}$ there are two regimes:
\begin{enumerate}
\item For the positive upper sign, initially the universe has a
subluminal expansion, followed by a superluminal expansion; \item
For the negative lower sign, the universe has an accelerated phase
during all its evolution.
\end{enumerate}
\end{enumerate}
Is the case (2.a) that it will interest us. In such a case, the
interpolation between a decelerated phase and an accelerated phase
can be achieved. However, there is high price to pay in order to
obtain such a simple qualitative realization of the accelerated
expansion today without introducing any kind of exotic matter: the
gravitational coupling is negative. In fact, the gravitational
coupling is related to the scalar field $\phi$ by the relation
\cite{weinberg}
\begin{equation}
G = \frac{4 + 2\omega}{4 + 3\omega}\frac{1}{\phi} \quad .
\end{equation}
Hence, for $- 3/2 < \omega < - 4/3$, $G < 0$.
\par
The most popular model to explain the current acceleration of the
universe is the $\Lambda CDM$ model. The matter content is a
pressureless fluid and the cosmological constant. For a spatial
flat case, the Einstein's equation reduce to
\begin{equation}
3\biggr(\frac{\dot a}{a}\biggl)^2 = 8\pi G\rho_m + \Lambda \quad ,
\end{equation}
where $\rho_m = \rho_{m0}a^{-3}$.
This equation admits the solution
\begin{equation}
a = \biggr(\frac{3M}{4\Lambda}\biggl)^\frac{1}{3}\sinh^\frac{2}{3}\sqrt{\frac{3\Lambda}{4M}}t
\quad .
\end{equation}
where $M = 8\pi G\rho_{m0}/3$. The asymptotical behaviour are
\begin{eqnarray}
t \rightarrow 0 \quad &\Rightarrow& \quad  a \propto t^\frac{2}{3} \quad , \\
t \rightarrow \infty \quad  &\Rightarrow& \quad  a \propto e^{\sqrt{\frac{3\Lambda}{4M}}t} \quad .
\end{eqnarray}
As expected, initially the scale factor describes a dust dominated
universe, and later a cosmological constant dominated universe.

\section{Comparing with the supernovae type Ia data}

We intend now to compare the background model for the flat
pressureless fluid Brans-Dicke cosmology with the supernovae type
Ia data. This is made through the computation of the luminosity
distance function, given by
\begin{equation}
D_L = \frac{a_0}{a}r,
\end{equation}
where $a_0$ is the value of the scale factor today, $a$ is the
value of the scale factor at the time of the emission of the
radiation, and $r$ is the co-moving radial distance. From now on,
we fix $a_0 = 1$. With this normalization choice, the scale factor
at a given moment $t$ is related with the redshift $z$ by
\begin{equation}
a = \frac{1}{1 + z} \quad .
\end{equation}
Hence, the luminosity distance can be expressed as \cite{weinberg}
\begin{equation}
D_L = (1 + z)r \quad .
\end{equation}
\par
Usually, the observational data are expressed in terms of the
distance moduli given by
\begin{equation}
\label{mu0}
 \mu_0 = 5\log\biggr(\frac{D_L}{Mpc}\biggl) + 25 \quad .
\end{equation}
In general the task now would consist in using the Friedmann
equation as
\begin{equation}
\biggr(\frac{\dot a}{a}\biggl)^2 = \frac{8\pi
G}{3}\sum_{i=1}^n\rho_i \quad,
\end{equation}
where $\rho_i$ are the different matter components. If each of
this component obeys a separate conservation equation, it can
express as $\rho_i = \rho_{i0}a^{-3(1 + \alpha_i)}$, where
$\alpha_i$ is the barotropic equation of state (supposed to be
constant) defined by $p_i = \alpha_i\rho_i$. Using then the
relation between the scale factor and $z$, the luminosity function
can be expressed as an integral over $z$ which depends on the
density parameter today for each fluid. This procedure is
detailed, for example, in reference \cite{colistete3}. It must be
remarked also that the expression (\ref{mu0}) must be modified for
the case the gravitational coupling varies with $z$, as reported
in references \cite{amendola1,berro,uzan}. However, in the case we
consider here, the local physics is supposed to be given by
general relativity, the Brans-Dicke modifications intervening at
cosmological scales only. Hence, with this assumption, the
expression (\ref{mu0}) may be still used safely.
\par
For the Brans-Dicke cosmological model presented in the previous
section the procedure outlined above becomes impossible to be
applied due to the non-minimal coupling between gravity and the
scalar field. But, we can use the relation
\begin{equation}
r = c\int_{t_e}^{t_0}\frac{dt}{a(t)} \quad ,
\end{equation}
where $t_e$ is the emission time for the source at the redshift
$z$ and $t_0$ is the present time, together with the normalization
condition $a(t_0) = a_0 = 1$. The expression (\ref{g1}) may be
rewritten as
\begin{eqnarray}
a(x) = \biggr(\frac{x - \frac{t_+}{t_0}}{1 -
\frac{t_+}{t_0}}\biggl)^{r + s}x^{r - s} \quad,\\
r = \frac{1 + \omega}{4 + 3\omega} \quad , \quad s = \frac{\sqrt{1
+ \frac{2}{3}\omega}}{4 + 3\omega} \quad , \quad x = \frac{t}{t_+}
\quad .
\end{eqnarray}
The present time may be determined through the definition for the
Hubble parameter
\begin{equation}
H = \frac{\dot a}{a} = \frac{2rt - (r - s)t_+}{(t - t_+)t} \quad .
\end{equation}
Evaluated for today, $t = t_0$, when $H = H_0$, this leads to the
expression
\begin{equation}
t_0 = \frac{(H_0t_+ + 2r) \pm \sqrt{(H_0t_+ + 2r)^2 - 4H_0(r -
s)t_+}}{2H_0} \quad ,
\end{equation}
where $H_0 = 100\,h\,km/(Mpc.s)$, $h$ being one of the parameters
to be determined. Hence, the luminosity distance is given by,
\begin{equation}
D_L = (1 + z)ct_0\int_x^1\frac{dy}{a(y)} \quad ,
\end{equation}
where $y$ is the integration variable. The same procedure is used
for the $\Lambda CDM$ model described in the end of the previous
section.
\par
In order to compare with the supernovae data, the (dimensionless)
time $x$ of emission must be determined from the expression for
the scale factor and the relation between the scale factor and the
redshift. In the case of the $\Lambda CDM$ model, there are two
parameters: $h$ and $\Lambda$ (or equivalently, $\rho_{m0}$). In
the Brans-Dicke model, there are in principle three free
parameters: $\omega$, $t_+$ and $h$. However, we first inspect the
best value for $t_+$, and then we vary the parameters $\omega$ and
$h$ around this parameter. The best fitting are obtained for $t_+
= 5.2\times10^{18}$: other values leads to an universe that
accelerates too early or too late. Hence, in both models, we are
left with two free parameters. We use the supernovae type Ia "gold
sample", with 157 high redshift supernovae, described in reference
\cite{riessbis} (see also reference \cite{colistete3}).
\begin{figure}[!htb]
\begin{minipage}[b]{0.32\linewidth}
\includegraphics[width=\linewidth]{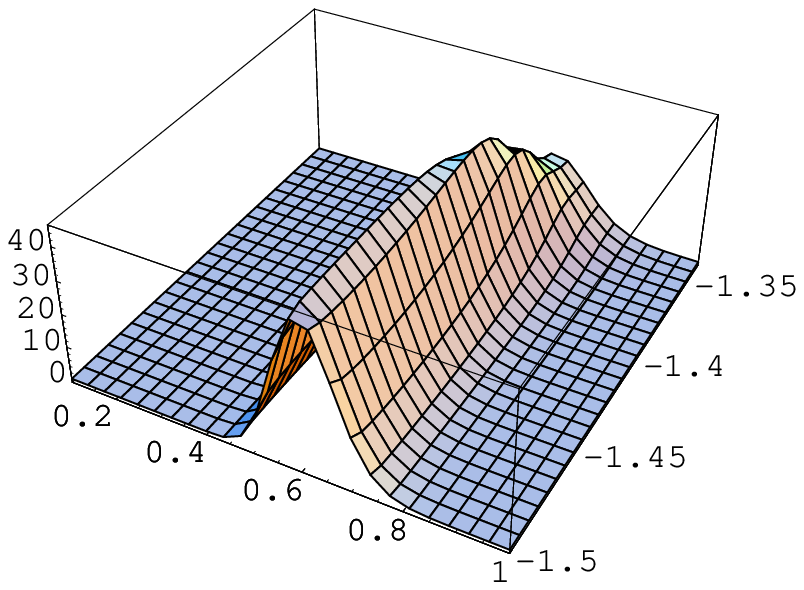}
\caption{{\protect\footnotesize Two-dimensional probability distribution for $\omega$ and
$h$.}}
\end{minipage} \hfill
\begin{minipage}[b]{0.32\linewidth}
\includegraphics[width=\linewidth]{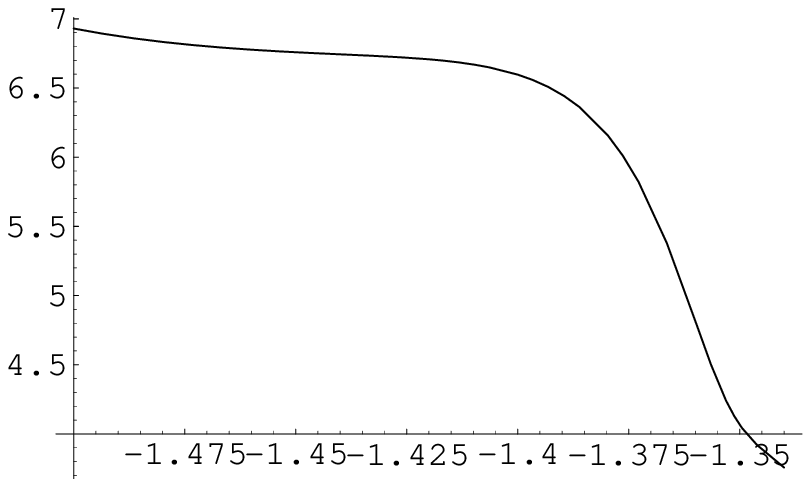}
\caption{{\protect\footnotesize Probability distribution for the parameter $\omega$.}}
\end{minipage} \hfill
\begin{minipage}[b]{0.32\linewidth}
\includegraphics[width=\linewidth]{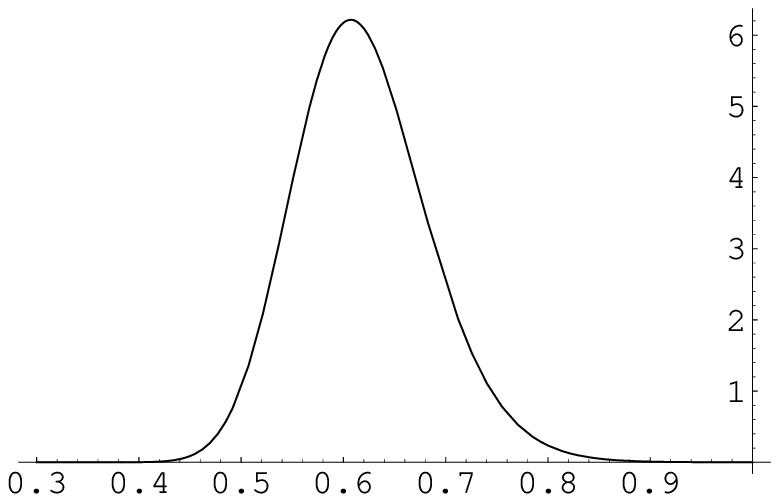}
\caption{{\protect\footnotesize Probability distribution for the parameter $h$.}}
\end{minipage} \hfill
\begin{minipage}[h]{0.32\linewidth}
\includegraphics[width=\linewidth]{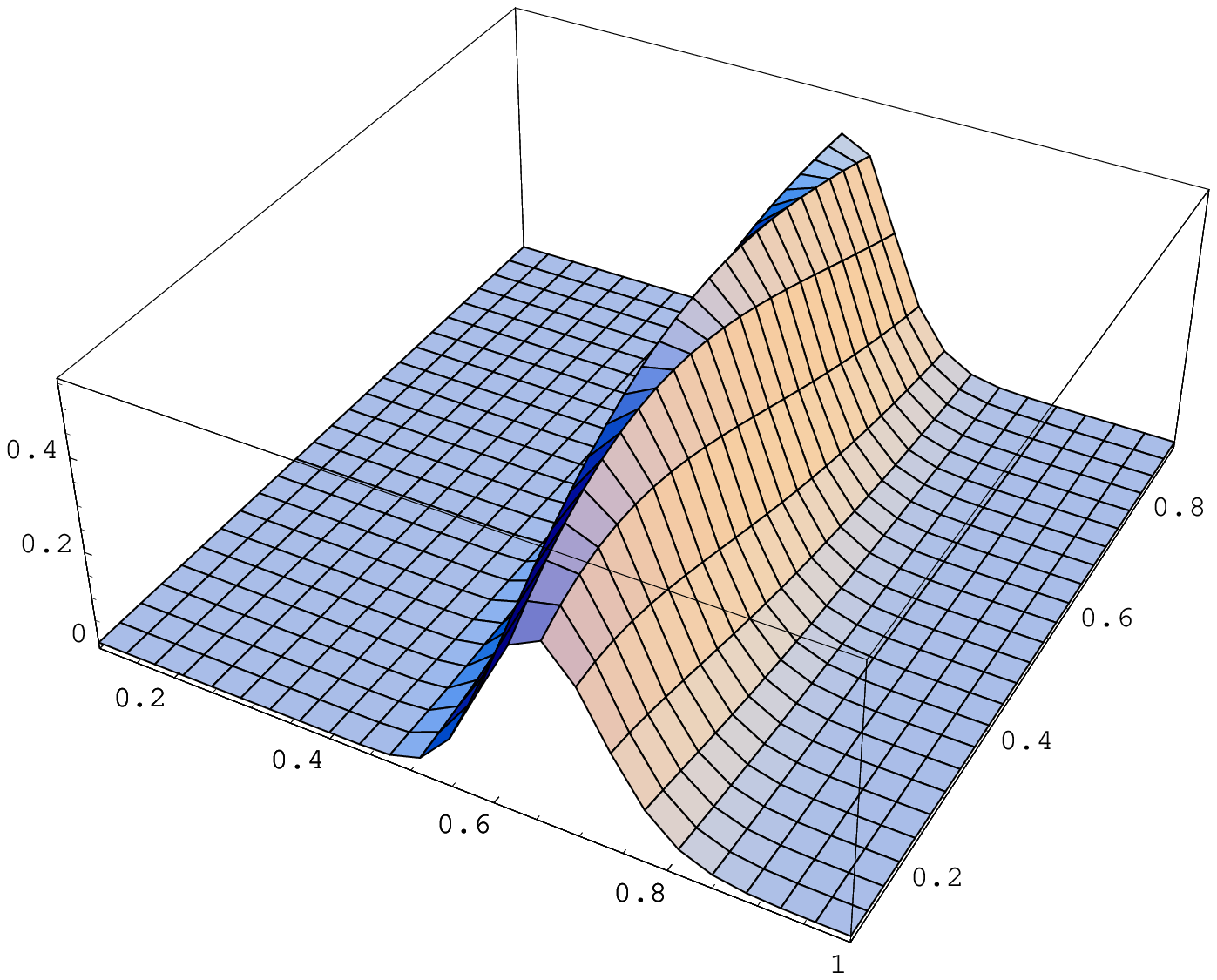}
\caption{{\protect\footnotesize Probability distribution for $\Omega_m$ and $h$ in the
$\Lambda CDM$ model.}}
\end{minipage} \hfill
\begin{minipage}[h]{0.32\linewidth}
\includegraphics[width=\linewidth]{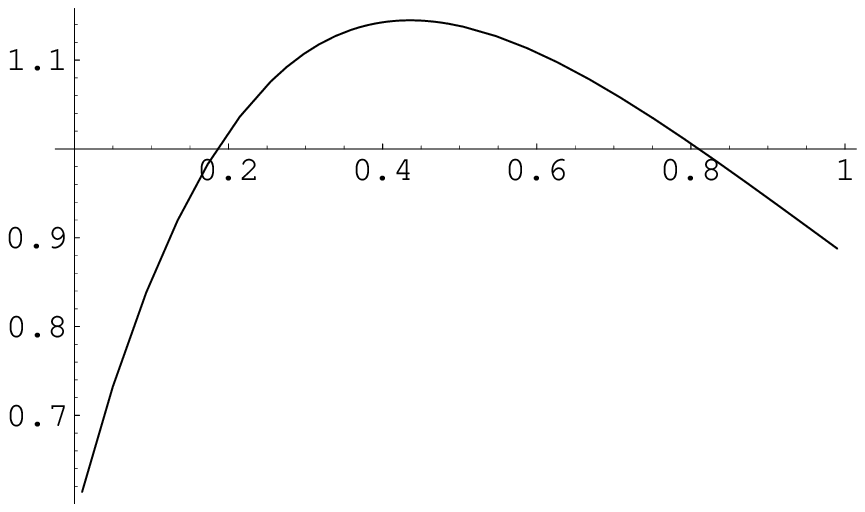}
\caption{{\protect\footnotesize Probability distribution for $\Omega_m$ in the $\Lambda
CDM$ model.}} \hfill
\end{minipage}
\begin{minipage}[h]{0.32\linewidth}
\includegraphics[width=\linewidth]{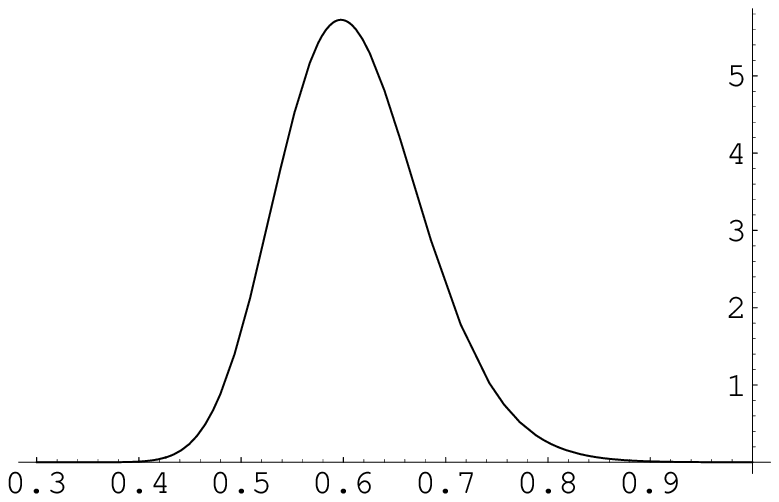}
\caption{{\protect\footnotesize Probability distribution for $h$ in the $\Lambda CDM$
model.}} \hfill
\end{minipage}
\end{figure}

\par
The main quantity to be evaluated is the $\chi^2$ function, which
gives the fitting quality:
\begin{equation}
\chi^2 = \sum_{i = 1}^n\frac{(\mu_{i0}^t -
\mu_{i0}^o)^2}{\sigma^2_i} \quad ,
\end{equation}
where $\mu_{i0}^t$ is the predicted theoretical value for the
$ith$ supernova, $\mu_{i0}^o$ its observational value and
$\sigma_i$ the error bar, taking already into account the effect
of peculiar dispersion. From this quantity, a probability function
is obtained:
\begin{equation}
P(\omega,h) = A\,e^{-\frac{\chi^2}{2}} \quad ,
\end{equation}
where $A$ is a normalization constant. The probability function
for a unique parameter can be obtained by marginalizing on the
remaining one:
\begin{equation}
P_\omega = \frac{\int P(\omega,h)dh}{\int\int P(\omega,h)dh\,d\omega}
\quad , \quad P_h = \frac{\int P(\omega,h)d\omega}{\int\int
P(\omega,h)dh\,d\omega} \quad .
\end{equation}
\par
In figures $1$, $2$ and $3$ the two dimensional probability
distribution for $\omega$ and $h$, as well as the one-dimensional
probability distribution for $\omega$ and $h$, in the Brans-Dicke
flat model, are displayed. The most probable value for $\omega$ is
$-1.5$, exactly the limiting case for the validity of solutions
(\ref{g1}). However, $\omega = - 1.5$ represents the case where
the Brans-Dicke theory is conformally equivalent to the Einstein's
theory. Hence, the prediction should more properly stated by
saying that the most probable value for the Brans-Dicke parameter
is $\omega \rightarrow  - 1.5$. For $h$, the most probable value
is $h = 0.605$. The best fitting for the Brans-Dicke flat model is
given by $\omega = -1.477$ and $h = 0.594$, with $\chi^2 =
1.19318$.
\par
In figures $4$, $5$ and $6$  the two-dimensional probability for
$\Omega_m$ and $h$ and the one-dimensional probability function
$\Omega_m$ and $h$, in the case of the $\Lambda CDM$ model, are
displayed. For $h$, the prediction is quite similar to the
Brans-Dicke case: the most probable value, marginalizing over
$\Omega_m$, is $h = 0.597$. The most probable value for the dark
matter density parameter, marginalizing over $h$, is $\Omega_m =
0.436$. The best fitting for the $\Lambda CDM$ model is given by
$\Omega_m = 0.506$ and $h = 0.597$ with $\chi^2 = 1.19813$. The
Brans-Dicke flat model gives a fitting slightly (but not
negligible) better than the $\Lambda CDM$ model.

\section{Conclusions}

The general solutions for the Brans-Dicke flat cosmological model,
determined by Gurevich et al, predict a late time accelerated
universe for $- 3/2 < \omega < - 4/3$, with an initial decelerated
phase. This solution can be a candidate to describe the observed
universe. In the present work, we have constrained this
Brans-Dicke flat model using the supernovae type Ia "gold sample".
We have compared the results with those obtained in a simplified
$\Lambda CDM$ model. The Brans-Dicke model leads to a slightly
smaller $\chi^2$ (which characterizes the quality of the fitting
of the observational data) with respect to the $\Lambda CDM$
model.
\par
In both models, the predicted value for the Hubble parameter $h$
is around $0.6$. This contrasts strongly with the estimation of
$h$ coming from the anisotropy of the cosmic microwave background
radiation, which leads to $h = 0.72\pm0.05$ \cite{spergel}.
However, this seems to be a general feature of the estimation of
$h$ using supernovae data and CMB data \cite{colistete3}. For the
Brans-Dicke model, the best value for the parameter $\omega$ is
around $- 1.5$. It must be remarked that the case $\omega = - 1.5$
implies that the Brans-Dicke theory is conformally equivalent to
general relativity. The minimum value for $\chi^2$ is obtained for
$h = 0.594$ and $\omega = - 1.477$, with $\chi^2 = 1.19318$. In
the $\Lambda CDM$ case, the best fitting is obtained with
$\Omega_m = 0.506$ and $h = 0.597$, with $\chi^2 = 1.19813$
\par
The results reported here indicate that, from the point of view of
the supernovae type Ia data, the Brans-Dicke cosmological flat
model leads to a quite viable scenario, even if the generalization
to the non-flat cases should be made in order to have a more
complete statistical analysis. However, at this point, this model
must be seen as essentially a toy model, mainly due to the
negative value for the parameter $\omega$. It would be interesting
to connect the Brans-Dicke model exploited here with effective
actions in four dimensions coming from M-theory and F-theory,
which predict also a negative value for the parameter $\omega$.
\par
The main problem with this scenario is that the values of $\omega$
are largely outside the values obtained with local tests, which
indicate $\omega > 1,000$ \cite{will}.  In reference \cite{orfeu},
indications for a negative $\omega$ have also been obtained.
Moreover, the range $- 3/2 < \omega < - 4/3$ implies a negative
gravitational coupling \cite{weinberg}. These drawbacks can be, in
principle, circumvented if the gravitational coupling is
scale-dependent, as suggested by considering quantum effects
\cite{reuter,shapiro}. In the present work, we made no attempt to
reconcile the values for $\omega$ deduced from local tests with
the corresponding values deduced from the analysis performed here.
But, evidently, this problem must be addressed in order to have a
complete realistic scenario.
\par
One important point to be signed is that previous study indicate
that structures can form in the Brans-Dicke model considered here
during all the evolution of the universe, after the radiative
phase, even the gravitational coupling is, at large scale,
repulsive \cite{fabris2,sergio}. However, a more detailed
comparison between the theoretical predictions for matter
agglomeration and the observational data must be made in order to
constraint more strongly the model. In special, the possibility
that the model can be unstable due to the repulsive gravitational
coupling must be verified \cite{brasil}. \vspace{0.5cm}
\newline\noindent {\bf Acknowledgments:} We thank J\'er\^ome
Martin for his remarks and criticisms. This work has been
partially supported by CNPq (Brazil).

\end{document}